\documentclass[a4paper]{jpconf}
\usepackage{lmodern}
\usepackage{amssymb,amsfonts,amsmath,,amscd}
\newcommand{\be}{\begin{equation}}
\newcommand{\ee}{\end{equation}}
\newcommand{\bea}{\begin{eqnarray}}
\newcommand{\eea}{\end{eqnarray}}
\newcommand{\nn}{\nonumber \\}
\newcommand{\p}[1]{(\ref{#1})}
\newcommand{\lb}{\label}

\begin{document}

\title{Off- and on-shell harmonic superspaces for \break
$6D$ SYM theories} \vspace{0.8cm}

\author{Evgeny Ivanov}
\vspace{0.4cm}

\address{Bogoliubov Laboratory of Theoretical Physics,
JINR,
141980, Dubna, Moscow Region, Russia}
\ead{eivanov@theor.jinr.ru}



\begin{abstract}
{It is a brief account of the harmonic superspace formulations of ${\cal N}=(1,0)$ and ${\cal N}=(1,1)$ SYM theories in six dimensions. The on-shell
${\cal N}=(1,1)$ harmonic superspace is argued to provide an efficient tool of constructing candidate counterterms and other invariants of ${\cal N}=(1,1)$ SYM.
It allows one, e.g., to find out an essential difference between the single- and double-trace dimension $\underline{d=10}$ invariants, which could be capable to explain
the absence of the three-loop double-trace (non-planar) counterterms in this theory. The defining superspace constraints of ${\cal N}=(1,1)$ SYM are solved in terms
of ${\cal N}=(1,0)$ harmonic superfields.}
\end{abstract}
\section{Motivations}
During recent years, much attention was paid to the maximally extended (with 16 supercharges) supersymmetric  gauge theories
in diverse dimensions (see, e.g., \cite{Seib}),
\bea
{\cal N}=4\,,\; 4D \quad \Longrightarrow \quad {\cal N}= (1,1)\,, \;6D \quad \Longrightarrow \quad {\cal N}= (1, 0)\,,\; 10D\,. \nonumber
\eea
The renowned ${\cal N}=4, 4D$ SYM theory was the first example of an UV {\it finite} theory.
Perhaps, it is also {\it completely integrable} \cite{ComplInt}.
The ${\cal N}=(1,1), 6D$ SYM is not renormalizable by formal counting (the coupling constant is dimensionful)
 but it is also expected to feature unique properties. In particular, in the perturbation theory it exhibits ``dual conformal symmetry'', like
 its $4D$ counterpart which respects ``dual superconformal symmetry'' \cite{DualConf}. It can be viewed as an effective theory for description
 of some particular low energy sectors of string theory, such as D5-brane dynamics. The quantum effective action of D5-brane as a generalization
 of the microscopic ${\cal N}=(1,1)$ SYM action
was conjectured to be of non-abelian Born-Infeld type \cite{Tseytlin1997,Drummond:2003ex}. The ${\cal N}=(1,1)$ SYM is anomaly free,
as opposed to ${\cal N}=(1,0)$ SYM theory.

The ${\cal N}=(1,1)$ and ${\cal N}=(1,0)$ SYM theories provide an appropriate ``laboratory'' for studying ${\cal N}=8$ supergravity
and its some lower ${\cal N}$ descendants, which are also non-renormalizable according by the standard counting.

The latest perturbative calculations in ${\cal N}=(1,1)$ SYM showed plenty of unexpected cancelations
of the UV divergencies. The theory is  UV finite up to 2 loops, while at 3 loops only a single-trace (planar) counterterm
of canonical dim 10 is needed. The permissible double-trace (non-planar) counterterms do not appear \cite{Bern:2005iz} - \cite{Bern:2012uf}.
Various arguments to explain this phenomenon were suggested \cite{Bossard:2010pk} - \cite{Bjornsson:2010wu}, though the complete
understanding of it is still lacking. One could expect the existence of some new non-renormalization theorems in this connection.

The maximal off-shell supersymmetry  that one can achieve in $6D$  is ${\cal N}=(1,0)$ supersymmetry. The natural off-shell formulation of
${\cal N}=(1,0)$ SYM theory is given in harmonic ${\cal N}=(1,0), 6D$ superspace  \cite{HSW,Zup1}, generalizing the harmonic ${\cal N}=2, 4D$  one \cite{harm1,harm}.
The harmonic $6D$ formulations were further worked out
in \cite{ISZ} - \cite{BuPl2} and \cite{BIS15}. The ${\cal N}=(1,1)$ SYM theory in the harmonic formalism is a hybrid of two ${\cal N}=(1,0)$ theories,
$[ {\cal N}=(1,1)\,\, {\rm SYM} ]$ = $[{\cal N}=(1,0)\,\, {\rm SYM} ]$
+ $[6D$ hypermultiplets$]$, with the second hidden ${\cal N}=(0,1)$ supersymmetry. The natural question is as to how to construct higher-dimension
${\cal N}=(1,1)$ invariants in the ${\cal N}=(1,0)$ superfield approach.

One way to approach this issue is the ``brute-force'' method. One starts with the appropriate dimension ${\cal N}=(1,0)$ SYM invariant and then
completes it to ${\cal N}=(1,1)$ invariant step by step, adding the proper hypermultiplet terms. This approach
is rather cumbersome.

Some simplifications arise due to the fact that for finding superfield counterterms it suffices to stay on
the mass shell. In a recent paper \cite{BIS15} there was suggested a new approach
to constructing higher-dimension  ${\cal N}=(1,1)$ invariants. It makes use of  the concept of the {\it on-shell}
${\cal N}=(1,1)$ harmonic superspace with the double
set of the harmonic variables $u^\pm_i, u^{\hat{\pm}}_A, i= 1,2; A = 1,2$ \cite{Bossard:2009sy}. The novel point of the consideration in \cite{BIS15}
is solving the ${\cal N}=(1,1)$ SYM constraints \cite{HST,HoweStelle} in terms of ${\cal N}=(1,0)$ superfields.
The dimension $\underline{d=8}$ and $\underline{d=10}$ invariants were
constructed in a simple way and an essential difference between the single- and double-trace dimension $\underline{d=10}$ invariants was observed. The present
contribution is a brief survey of the $6D$ harmonic methods, with the main emphasis on their recent applications in \cite{BIS15}.

\setcounter{equation}{0}
\section{$6D$ superspaces and superfields}

\subsection{$6D$ superspaces}
\begin{itemize}
  \item
The customary ${\cal N}=(1,0), 6D$ superspace is parametrized by the following coordinate set:
\bea
z=(x^M, \theta^a_i)\,, \quad  M= 0, \ldots, 5\,, \; a = 1,\ldots, 4\,, \; i = 1, 2\,,
   \lb{6Dreal}
\eea
\item
The harmonic ${\cal N}=(1,0)$ superspace amounts to adding $SU(2)$ harmonics to \p{6Dreal}:
\bea
Z := (z, u) = (x^M, \theta^a_i, u^{\pm i})\,, \quad u^-_i = (u^+_i)^* , u^{+i} u_i^- = 1\,, \; u^{\pm i} \in SU(2)_R/U(1)\,.
    \lb{6Dharm}
\eea
\item
The {\it analytic} ${\cal N}=(1,0)$ superspace forms an invariant subspace in \p{6Dharm}:
\bea
\zeta :=(x^M_{({\rm an})}, \theta^{+a}, u^{\pm i}) \subset Z\,, \quad  x^M_{({\rm an})}=x^M+ \frac i2 \theta^a_k\gamma^M_{ab}\theta^b_l u^{+k}u^{-l}, \quad
\theta^{\pm a} =\theta^a_iu^{\pm i}\,.
\lb{analyt}
\eea
\end{itemize}
\noindent The differential operators in the analytic basis $Z_A := (x^M_{({\rm an})}, \theta^{+a}, u^{\pm i}, \theta^{-a})$ are defined according to
\bea
&& D^+_a=\partial_{-a}~, \;
D^-_a=-\partial_{+ a}-2i\theta^{-b}\partial_{ab}\,, \; D^0 = u^{+i} \frac {\partial}{ \partial u^{+i}} -
u^{-i} \frac {\partial}{ \partial u^{-i}} + \theta^{+a} \partial_{+ a} -  \theta^{-a} \partial_{- a} \nn
&& D^{++}=\partial^{++}+i\theta^{+a}\theta^{+b}\partial_{ab}+\theta^{+a}\partial_{-a}~,\; D^{--}=\partial^{--}+i\theta^{-a}
\theta^{-b}\partial_{ab}+\theta^{-a}\partial_{+a}~,\lb{derivat}\\
&&\partial_{\pm a}\theta^{\pm b} = \delta^b_a\,, \quad \partial^{++} =  u^{+i} \frac {\partial }{ \partial u^{-i}} , \quad
\partial^{--}=  u^{-i} \frac {\partial }{ \partial u^{+i}}\, .\nonumber
\eea

\subsection{$6D$ superfields}
The fundamental quantity of ${\cal N}=(1,0)$ SYM theory is the analytic gauge connection $V^{++}(\zeta)$
\bea
\nabla^{++}= D^{++}+ V^{++}\,, \quad \delta V^{++} = -\nabla^{++}\Lambda\,, \quad \Lambda = \Lambda(\zeta)\,.
\lb{Prep}
\eea
The second harmonic (non-analytic) connection $V^{--}(Z)$ entering the covariantized harmonic derivative $\nabla^{--}$,
$$
\nabla^{--} = D^{--} + V^{--}\,, \quad \delta V^{--} = -\nabla^{--}\Lambda\,,
$$
is expressed through $V^{++}$ from the harmonic ``flatness'' condition
\bea
&&[\nabla^{++}, \nabla^{--}] = D^0 \;\Leftrightarrow \; D^{++}V^{--} - D^{--}V^{++} +[V^{++}, V^{--}]=0 \nn
&&\Rightarrow\, V^{--} = V^{--}(V^{++}, u^\pm)\,.
\lb{flateness}
\eea

The off-shell contents of ${\cal N}=(1,0)$ SYM theory is revealed in the Wess-Zumino gauge for the analytic gauge potential:
\bea
V^{++} \ =\ \theta^{+a} \theta^{+b} A_{ab} + 2 (\theta^+)^3_a \lambda^{ai} u_i^- - 3 (\theta^+)^4 {\cal D}^{ik} u^-_i u^-_k \, .\lb{WZ}
 \eea
Here $A_{ab}$ is the gauge field, $\lambda^{ai}$ is the gaugino and ${\cal D}^{ik} = {\cal D}^{ki}$ are the auxiliary fields.

The ${\cal N}=(1,0)$ SYM covariant derivatives are defined as
\bea
&&{\nabla}^-_a = [\nabla^{--}, D^+_a] = D^-_a + {\cal A}^-_a , \ \ \ \ \ \ \nabla_{ab} = \frac{1}{2i}[D^+_a, \nabla^-_b] =
\partial_{ab} + {\cal A}_{ab} \,, \nn
&& {\cal A}^-_a(V)=-D^+_a V^{--},\quad {\cal A}_{ab}(V)= \frac{i}{2}D^+_aD^+_b V^{--}, \nn
&& [\nabla^{++}, \nabla^{-}_a] = D^+_a\,, \quad [\nabla^{++}, D^{+}_a] = [\nabla^{--}, \nabla^{-}_a] = [\nabla^{\pm\pm}, \nabla_{ab}] = 0\,.
\lb{CovDer}
\eea

The covariant superfield strengths are constructed by the appropriate connections
\bea
&& [D^+_a, \nabla_{bc}] \ =\ \frac{i}2\varepsilon_{abcd} W^{+d} \, , \ \ \ \ \ \ \
[\nabla^-_a, \nabla_{bc}] \ =\ \frac{i}{2}\varepsilon_{abcd} W^{-d}  \,,\nn
&&  W^{+a}=-\frac{1}{6}\varepsilon^{abcd}D^+_b D^+_c D^+_d V^{--}\,,\quad W^{- a} := \nabla^{--}W^{+a}\,, \nn
&& \nabla^{++} W^{+a}  = \nabla^{--} W^{-a} \ =\ 0\,,
\quad \nabla^{++}W^{-a} = W^{+ a}\,,  \nn
&& D^+_b W^{+ a} = \delta^a_b F^{++}\,, \quad  F^{++}=\frac14 D^+_aW^{+a}=(D^+)^4 V^{--}\, ,\; \nn
&& \nabla^{++}F^{++} = 0\,, \quad D^+_a F^{++} = 0\,.\lb{CovStr}
\eea

The hypermultiplet is described off shell by an analytic superfield $q^{+ A}(\zeta)\,, ( A = 1,2),$ :
\bea
q^{+ A}(\zeta) = q^{iA}(x) u^+_i - \theta^{+ a}\psi^A_a(x) + \; \rm{An \;infinite \; tail \; of \;auxiliary \;fields}\,.\lb{Hyper}
\eea

\subsection{${\cal N}=(1,0)$ superfield actions}
The ${\cal N}=(1,0)$ SYM action was constructed by Boris Zupnik \cite{Zup1}:
\bea
&& S^{SYM} =\frac{1}{f^2}\sum\limits^{\infty}_{n=1} \frac{(-1)^{n+1}}{n} {\rm Tr} \int
d^6\!x\, d^8\theta\, du_1\ldots du_n \frac{V^{++}(z,u_1 )
\ldots V^{++}(z,u_n ) }{(u^+_1 u^+_2)\ldots (u^+_n u^+_1 )}\,,\nn
&& \delta S^{SYM} = 0 \; \Rightarrow \; F^{++} = 0\,.
\lb{SymAct}
\eea
Here, $(u^+_1 u^+_2)^{-1}, \ldots (u^+_n u^+_1 )^{-1} $ are harmonic distributions \cite{harm}.

The hypermultiplet action, with $q^{+A}$ in adjoint representation of the gauge group, is written down as
\bea
&& S^q = -\frac{1}{2 f^2}{\rm Tr} \int d\zeta^{-4} q^{+ A}\nabla^{++} q^+_A\,, \quad \nabla^{++} q^+_A = D^{++}q^+_A + [V^{++}, q^+_A]\,, \nn
&& \delta S^q = 0 \; \Rightarrow \; \nabla^{++} q^{+ A} = 0\,.
\lb{HypAct}
\eea

The ${\cal N}=(1,0)$ superfield form of the ${\cal N}=(1,1)$ SYM action is a sum of the two superfield actions given above:
\bea
&& S^{(V + q)}= S^{SYM} + S^q = \frac{1}{f^2}\left(\int dZ {\cal L}^{\rm SYM} - \frac12 {\rm Tr}\int d\zeta^{-4} q^{+ A}\nabla^{++} q^+_A \right), \nn
&& \delta S^{(V + q)} = 0 \; \Rightarrow \;F^{++} + \frac12 [q^{+ A}, q_A^+] = 0\,, \quad \nabla^{++} q^{+ A} = 0\,.
\lb{HybrAct}
\eea
It is invariant under the second hidden ${\cal N}= (0, 1)$ supersymmetry acting as:
\bea
\delta V^{++} = \epsilon^{+ A}q^+_A\,, \quad \delta q^{+ A} = -(D^+)^4 (\epsilon^-_A V^{--})\,, \quad
\epsilon^{\pm}_A = \epsilon_{aA}\theta^{\pm a}\,.\lb{Hidden}
\eea
These transformations have the correct closure among themselves and with the manifest ${\cal N}=(1,0)$ supersymmetry
only on shell.

\setcounter{equation}{0}
\section{Higher-dimensional invariants}

\subsection{ Dimension $\underline{d=6}$}

In the pure ${\cal N}=(1,0)$ SYM theory the $\underline{d=6}$ invariant is defined uniquely \cite{ISZ}:
\bea
S_{SYM}^{(6)}  = \frac{1}{2g^2}{\rm Tr}\int  d\zeta^{-4}du \,\left(F^{++}\right)^2 \;\sim \;  {\rm Tr}\int d^6x [ (\nabla^M F_{ML})^2 + \ldots].
\lb{dim6}
\eea
It vanishes on shell, when $F^{++}=0\,$. Using the results of \cite{IShyper}, its ${\cal N}=(1,1)$ completion is defined up to a real parameter
\bea
{\cal L}^{d=6} = \frac{1}{2g^2} {\rm Tr} \int du d\zeta^{-4}\, \left( F^{++} + \frac12 [ q^{+ A}, q^+_A] \right) \left(
F^{++} + 2\beta [ q^{+ A}, q^+_A] \right).\lb{dim62}
\eea
But it vanishes by itself on the full ${\cal N}=(1,1)$ SYM mass shell! This proves the {\it one-loop finiteness} of ${\cal N}=(1,1)$ SYM theory.

\subsection{Dimension $\underline{d=8}$}

All superfield operators of the canonical dimension $\underline{d=8}$ in the ${\cal N}=(1,0)$ SYM theory
vanish on shell, in accord with the statement of ref. \cite{HoweStelle}. Can this conclusion be changed upon adding the hypermultiplet terms?
We have checked that there exist no ${\cal N} = (1,0)$ off-shell invariants of the dimension $\underline{d=8}$ which would
respect the on-shell ${\cal N} = (1,1)$ invariance.

Surprisingly, the $\underline{d=8}$ superfield expression which is non-vanishing on shell and respects the on-shell
${\cal N}=(1,1)$ supersymmetry can be constructed by {\it giving up} the demand of {\it off-shell} ${\cal N}=(1,0)$ supersymmetry.

An example of such an invariant in ${\cal N}=(1,0)$ SYM theory is very simple
\bea
{\tilde S}^{(8)}_{1}  \sim  {\rm Tr} \int d\zeta^{-4}\,
\varepsilon_{abcd}W^{+ a} W^{+ b} W^{+ c} W^{+ d}\,. \lb{d810}
\eea
Indeed, $D^+_a W^{+b} = \delta_a^b F^{++}$, which vanishes on shell, with $ F^{++} =0\,$. Thus, $W^{+a}$ is on-shell analytic,
for which reason the above action respects ${\cal N}=(1,0)$ supersymmetry on shell. Also, an analogous double-trace on-shell invariant exists.

These invariants possess ${\cal N}=(1,1)$ completions. For \p{d810} such a completion reads
\bea
&& {\cal L}^{+4}_{(1,1)}  = {\rm Tr}_{(S)}\Big\{  \frac{1}{4} \varepsilon_{abcd}
 W^{+a} W^{+b} W^{+c} W^{+d}  + 3i q^{+A} \nabla_{ab} q_A^+ W^{+a} W^{+b} \nn
&&-\,  q^{+A} \nabla_{ab} q_A^+ \, q^{+B} \nabla^{ab} q_B^+
-  W^{+a} [ D_a^+ q_A^- , q_B^+] q^{+A} q^{+B} \nn
&& -\,\frac{1}{2} [ q^{+C} , q^+_C] [ q_A^-,q^+_B] q^{+A} q^{+B}\Big\}.
   \lb{dim81}
\eea
Here, ${\rm Tr}_{(S)}$ stands for the {\it symmetrized} trace. This Lagrangian is analytic, $D^+_a{\cal L}^{+4}_{(1,1)}=0$,
on the total mass shell $F^{++} + \frac12 [q^{+ A}, q^+_A] = 0$, $\nabla^{++}q^{+ A} =0\,$, hence it is on-shell ${\cal N}=(1,1)$ supersymmetric.

Though the nontrivial on-shell $\underline{d=8}$ invariants exist, the perturbative
expansion for the amplitudes in the ${\cal N}=(1,1)$ SYM theory  does not involve divergences at the two-loop level.
The reason is that these $\underline{d=8}$ invariants do {\it not} possess the full off-shell ${\cal N} = (1,0)$ supersymmetry
 which the physically relevant counterterms should obey.

\setcounter{equation}{0}
\section{${\cal N}=(1,1)$ on-shell harmonic superspace}

Apart from the fact that  the $\underline{d=8}$ terms mentioned above cannot come out as counterterms in ${\cal N}=(1,1)$ SYM theory,
they can arise, e.g.,
as quantum corrections to the effective Wilsonian action. For the pure ${\cal N}=(1,0)$ SYM theory this was recently
observed in \cite{BuPl2}.  It was desirable to have some systematic way of constructing
such higher-order ${\cal N}=(1,1)$ invariants. This proves to be possible within
the on-shell harmonic ${\cal N}=(1,1)$ superspace.

Let us start by extending the ${\cal N}=(1,0)$ superspace to the ${\cal N}=(1,1)$ one,
\bea
z = (x^{ab}, \theta^a_i) \; \Rightarrow \; \hat{z} = (x^{ab}, \theta^a_i, \hat{\theta}^A_a)\,.
\lb{116D}
\eea
Then we introduce the gauge-covariantized spinor derivatives,
\bea
\nabla^i_a = \frac \partial {\partial \theta^a_i} - i \theta^{bi} \partial_{ab} + {\cal A}^i_a \, ,\quad
\hat{\nabla}^{aA}  = \frac \partial {\partial \hat{\theta}_{Aa}} - i \hat{\theta}_b^A \partial^{ab}  +
\hat{{\cal A}}^{aA}\,. \lb{Der11}
\eea
The superspace  constraints defining the ${\cal N}=(1,1)$ SYM theory can be then written as follows \cite{HST,HoweStelle}:
\bea
&& \{\nabla^{(i}_a, \nabla^{j)}_b \} = \{\hat{\nabla}^{a(A}, \hat{\nabla}^{bB)} \} = 0 \,, \quad \{\nabla^i_a, \hat{\nabla}^{bA} \} =
\delta_a^b \phi^{iA} \nn
&&  \Rightarrow \quad \nabla^{(i}_a \phi^{j)A}  =   \hat{\nabla}^{a(A}   \phi^{B)i}  = 0 \quad ({\rm By \; Bianchis})\,.\lb{Constr2}
 \eea

As the next step, we introduce the ${\cal N}=(1,1)$ harmonic superspace \cite{Bossard:2009sy},
\bea
Z = (x^{ab}, \theta^a_i, u^\pm_k) \; \Rightarrow \; \hat{Z} = (x^{ab}, \theta^a_i, \hat{\theta}^A_b, u^\pm_k, u^{\hat{\pm}}_A)\,,
\lb{BiHarm}
\eea
pass to the analytic basis in it and choose the ``hatted'' spinor derivatives short,
$\nabla^{\hat{+}a} = D^{\hat{+}a} = \frac{\partial}{\partial \theta_a^{\hat{-}}}\,.$ The set of constraints \p{Constr2} is equivalently rewritten as
\bea
&& \{\nabla^+_a, \nabla^+_{b} \}  = 0\,, \quad \{D^{\hat{+}a},
D^{\hat{+}b} \}  = 0\,, \quad \{\nabla^+_a, D^{\hat{+}b} \}  = \delta_a^b \phi^{+\hat{+}}\,, \nn
&& [ \nabla^{\hat{+}\hat{+}}, \nabla^+_a]  = 0\,, \; [\tilde{\nabla}^{{+}{+}}, \nabla^+_a]  = 0\,, \;
[\nabla^{\hat{+}\hat{+}}, D^{a\hat{+}}]  = 0\,, \;
[ \tilde{\nabla}^{++}, D^{a\hat{+}}]  = 0\,, \nn
&& [\tilde{\nabla}^{++}, \nabla^{\hat{+}\hat{+}}]  = 0\,, \lb{11HSSconstr} \\
&&\nabla^+_a = D^+_a + {\cal A}^+_a(\hat{Z})\,, \; \tilde{\nabla}^{++} = D^{++} + \tilde{V}^{++}(\hat{\zeta})\,, \quad \nabla^{\hat{+}\hat{+}} =
D^{\hat{+}\hat{+}} + V^{\hat{+}\hat{+}}(\hat{\zeta})\,,\nonumber \\
&&\hat{\zeta} = (x^{ab}_{\rm an}, \theta^{\pm a}, \theta^{\hat{+}}_c, u^{\pm}_i, u^{\hat{\pm}}_A)\,.\lb{11Der}
\eea

\setcounter{equation}{0}
\section{Solving ${\cal N}=(1,1)$ SYM constraints through ${\cal N}=(1,0)$ superfields}
The starting point of our analysis in \cite{BIS15} was the WZ gauge for
the extra connection $V^{\hat{+}\hat{+}}(\hat{\zeta})$
\bea
V^{\hat{+}\hat{+}} = i\theta^{\hat{+}}_a \theta^{\hat{+}}_b\hat{\cal A}^{ab} +
\varepsilon^{abcd}\theta^{\hat{+}}_a\theta^{\hat{+}}_b\theta^{\hat{+}}_c\varphi_d^{A}u^{\hat{-}}_A
+ \varepsilon^{abcd}\theta^{\hat{+}}_a\theta^{\hat{+}}_b\theta^{\hat{+}}_c\theta^{\hat{+}}_d{\cal D}^{AB}u^{\hat{-}}_Au^{\hat{-}}_B\,,
   \lb{WZ2}
\eea
where $\hat{\cal A}^{ab}, \varphi_d^A$ and ${\cal D}^{(AB)}$
are some ${\cal N}=(1,0)$ harmonic superfields, still arbitrary at this step.

Then the above constraints are reduced to some sets of harmonic equations.
We have solved them and, as the eventual result, found that the first harmonic connection $V^{++}$ coincides
with the previous ${\cal N}= (1,0)$ one, $ V^{++} = V^{++}(\zeta)$,
while the dependence of all other geometric ${\cal N}= (1,1)$ objects on the ``hatted'' variables is fixed as
\bea
&& V^{\hat{+}\hat{+}} =  i \theta_a^{\hat{+}} \theta_b^{\hat{+}} {\cal A}^{ab} -
 \frac 13 \epsilon^{abcd}
 \theta^{\hat{+}}_a \theta^{\hat{+}}_b
\theta^{\hat{+}}_c D^+_d q^{- \hat{-}}  + \frac 18  \epsilon^{abcd}
 \theta^{\hat{+}}_a \theta^{\hat{+}}_b
\theta^{\hat{+}}_c  \theta^{\hat{+}}_d [q^{+\hat{-}}, q^{-\hat{-}} ] \nn
&& \phi^{+\hat{+}} = q^{+ \hat{+}} - \theta^{\hat{+}}_a W^{+a} - i
 \theta^{\hat{+}}_a \theta^{\hat{+}}_b \nabla^{ab} q^{+\hat{-}}  + \frac{1}{6}  \varepsilon^{abcd}
 \theta^{\hat{+}}_a \theta^{\hat{+}}_b  \theta^{\hat{+}}_c [ D^+_d q^{-\hat{-}}, q^{+\hat{-}}] \nonumber \\
&& +\,  \frac{1}{24}  \varepsilon^{abcd}  \theta^{\hat{+}}_a \theta^{\hat{+}}_b
 \theta^{\hat{+}}_c\theta^{\hat{+}}_d  [ q^{+\hat{-}}, [ q^{+\hat{-}} , q^{-\hat{-}}] ] \,.
\lb{Vsec}
\eea
Here, $q^{+ \hat{\pm}} = q^{+ A}(\zeta)u^{\hat{\pm}}_A\,, \; q^{- \hat{\pm}} = q^{- A}(\zeta)u^{\hat{\pm}}_A$ and
$ W^{+ a}, q^{\pm A}$ are just the ${\cal N}=(1,0)$ superfields we dealt with previously.
In the course of solving the constraints, there naturally appear the superfield equations of motion
\bea
 \nabla^{++}q^{+ A} = 0\,, \quad F^{++} = \frac14 D^+_aW^{+ a} = - \frac12 [q^{+ A},q^+_A]\,.
 \lb{Eqfull}
\eea
Also, the structure of the spinor covariant derivatives is completely fixed
\bea
&& \nabla^+_a = D^+_a - \theta^{\hat{+}}_a q^{+\hat{-}} +  \theta^{\hat{-}}_a \phi^{+\hat{+}} \, ,\nn
&&\nabla^-_a = D^-_a - D^+_a V^{--} - \theta^{\hat{+}}_a q^{-\hat{-}} +  \theta^{\hat{-}}_a \phi^{-\hat{+}}\,, \quad
\phi^{-\hat{+}} = \nabla^{--}\phi^{+\hat{+}}\,.
 \lb{11Der2}
\eea

The crucial point of our analysis was the requirement that the vector $6D$ connections in the sectors of hatted
and unhatted variables are identical to each other.

The basic advantage of using the constrained ${\cal N}=(1,1)$ strengths  $\phi^{\pm\hat{+}}$
for constructing invariants is the very simple transformation rules of $\phi^{\pm\hat{+}}$ under the hidden ${\cal N}=(0,1)$
supersymmetry
\bea
\delta \phi^{\pm\hat{+}} = - \epsilon^{\hat{+}}_a \frac{ \partial\, }
{\partial \theta_a^{\hat{+}}} \phi^{\pm\hat{+}} - 2 i \epsilon_a^{\hat{-}}
 \theta_b^{\hat{+}} \partial^{ab} \phi^{\pm\hat{+}}  - [\Lambda^{(comp)}, \phi^{\pm\hat{+}}]\,,
\lb{01Symm}
\eea
where $\Lambda^{(comp)}$ is some composite gauge parameter which makes no contribution under the trace.

\setcounter{equation}{0}
\section{Invariants in ${\cal N}=(1,1)$ superspace}
The single-trace dimension $\underline{d=8}$ invariant \p{dim81} can be readily
rewritten in  ${\cal N}=(1,1)$ superspace
\bea
S_{(1,1)} = \int d\zeta^{-4}{\cal L}^{+4}_{(1,1)}\,, \; {\cal L}^{+4}_{(1,1)}   =
-{\rm Tr}\,\frac{1}{4}\int d\hat{\zeta}^{-4} d\hat{u} \,
(\phi^{+\hat{+}} )^4, \quad d\hat{\zeta}^{-4} \sim (D^{\hat{-}})^4
\lb{HSSsingle}
\eea
$$\delta {\cal L}^{+4}_{(1,1)} =  - 2 i \partial^{ab} {\rm Tr}  \int d\hat{\zeta}^{-4} d\hat{u} \,
\Big[ \epsilon_a^{\hat{-}} \theta_b^{\hat{+}}    \frac{1}{4} (\phi^{+\hat{+}})^4\Big]. $$
Analogously, the double-trace $\underline{d=8}$ invariant is given by
\bea
\hat{\cal L}^{+4}_{(1,1)} = -\frac14 \int  d\hat{\zeta}^{-4} d\hat{u} \,
{\rm Tr}\,(\phi^{+\hat{+}} )^2\, {\rm Tr}\,(\phi^{+\hat{+}} )^2.
\lb{HSSdouble}
\eea

Now it is easy to construct the single- and double-trace $\underline{d=10}$ invariants
\bea
&& S_1^{(10)} = {\rm Tr} \int  dZ d\hat{\zeta}^{-4} d\hat{u} \,
 (\phi^{+\hat{+}} )^2 (\phi^{-\hat{+}} )^2, \quad \phi^{-\hat{+}} =\nabla^{--}\phi^{+\hat{+}}\,, \nn
&& S_2^{(10)} = -\int dZ d\hat{\zeta}^{-4} d\hat{u} \,
{\rm Tr} \Big(\phi^{+\hat{+}}  \phi^{-\hat{+}}\Big)\,{\rm Tr} \Big(\phi^{+\hat{+}}  \phi^{-\hat{+}}\Big)\,.
\lb{2Dim10}
\eea

It is notable that the single-trace $\underline{d=10}$ invariant admits a representation as an integral over the {\it full}
${\cal N}=(1,1)$ superspace
\bea
S_1^{(10)}
\sim {\rm Tr} \int dZ d\hat{Z} d\hat{u}\ \phi^{+ \hat{+}} \phi^{- \hat{-}}\,, \quad \phi^{- \hat{-}} = \nabla^{\hat{-}\hat{-}} \phi^{- \hat{+}}\,,
    \lb{SecRepr}
\eea
with $d\hat{Z} \sim (D^{\hat{-}})^4(D^{\hat{+}})^4\,$. On the other hand, the double-trace $\underline{d=10}$ invariant {\it cannot} be written
as the total integral and so looks as being UV {\it protected}.

This could explain why in the perturbative calculations of the amplitudes in the ${\cal N}=(1,1)$ SYM theory
single-trace 3-loop divergence is seen, while no double-trace structures at the same order were observed \cite{Bern:2005iz}, \cite{Bern:2010tq},
\cite{Bern:2012uf}. However, this does not seem to be like the standard non-renormalization theorems because the quantum calculation of
${\cal N}=(1,0)$ supergraphs should give invariants in the {\it off-shell} ${\cal N}=(1,0)$ superspace, not
in the {\it on-shell} ${\cal N}=(1,1)$ superspace. So the above property seems not enough to explain the absence
of the double-trace divergences and some additional piece of reasoning  is needed.

\section{Summary and outlook}

Based on refs. \cite{Zup1}, \cite{ISZ}, \cite{IShyper} and \cite{BIS15}, the off-shell ${\cal N}=(1,0)$
and on-shell  ${\cal N}=(1,1)$ harmonic superfield
approaches were accounted for. It was argued that they are very efficient for constructing higher-dimensional invariants in the ${\cal N}=(1,0)$ and  ${\cal N}=(1,1)$ SYM
theories. The novel solution of the ${\cal N}=(1,1)$ SYM constraints in terms of
 harmonic ${\cal N}=(1,0)$ superfields was given.  This allowed us to explicitly construct the full set of
 the dimensions $\underline{d=8}$ and $\underline{d=10}$  superfield invariants revealing ${\cal N}=(1,1)$
 on-shell supersymmetry.

All possible $\underline{d=6}$ ${\cal N}=(1,1)$ invariants were shown to be  on-shell vanishing,
thus proving the {UV} finiteness of ${\cal N}=(1,1)$ SYM at one loop.

The off-shell $\underline{d=8}$ invariants which would be non-vanishing on shell, are absent. Assuming that the ${\cal N}=(1,0)$ supergraphs
yield integrals over the full ${\cal N}=(1,0)$ harmonic superspace, this means the absence
of two-loop counterterms as well.

Two dimension $\underline{d=10}$ invariants were constructed as integrals over the whole ${\cal N}=(1,0)$ harmonic superspace.
The single-trace invariant can be rewritten as an integral over ${\cal N}=(1,1)$ superspace,
while the double-trace one cannot. This property combined with an additional reasoning (e.g., based on the algebraic renormalization
scheme \cite{Piguet:1995er}) could explain why the double-trace invariant is {UV} protected.
\vspace{0.3cm}

\noindent \underline{\it Some further lines of development}:
 \vspace{0.3cm}

\noindent{\bf (a).} It would be tempting and instructive to construct the $\underline{d\geq 12}$ invariants in the ${\cal N}=(1,1)$
SYM theory using the on-shell ${\cal N}=(1,1)$ harmonic superspace techniques and to see whether they exhibit the properties similar to the
$\underline{d=10}$ invariants. It is probable that the proper corrections to the hidden supersymmetry transformations will
be of need, while tackling this issue.

\noindent{\bf (b).} It is worth to apply the same method for constructing the Born-Infeld action with the manifest off-shell ${\cal N}=(1,0)$
and hidden on-shell ${\cal N}=(0,1)$ supersymmetries.

\noindent{\bf (c).} The closely related problem is to recover the higher dimension invariants listed above from the quantum ${\cal N}=(1,0)$
supergraph techniques. The first steps in this direction were undertaken in a recent paper \cite{BIMS}.

\noindent{\bf (d).} An interesting task is to develop an analogous on-shell bi-harmonic ${\cal N} =4, 4D$ superspace approach to
 the ${\cal N} =4, 4D$ SYM theory in the  ${\cal N}=2$ superfield formulation (by solving the ${\cal N} =4$ SYM constraints
 in terms of ${\cal N}=2$ superfields) and apply it to the problem of constructing the ${\cal N} =4$ SYM effective action.
 It is curious that such a formulation has not been constructed so far in full generality, despite the existence of various more sophisticated
superspace formulations (see \cite{BIS} for a recent review).

\noindent{\bf (e).} As was mentioned in sect. 5, the crucial last step in solving the constraints of ${\cal N}=(1,1)$ SYM theory was
identifying the vector connections in the sectors with the standard and ``hatted'' harmonics. Only after this identification, the constraints in the bi-harmonic superspace
get fully equivalent to those in the standard setting and imply the equations of motion for the involved superfields. It would be interesting
to develop a superspace formulation with two {\it independent} vector connections and to inquire whether it could give rise to an off-shell description of ${\cal N}=(1,1), 6D$
(and, perhaps, of ${\cal N}=4, 4D$) SYM theories.  The introduction of two vector connections seems to imply doubling of the $x$-coordinate, in an obvious parallel
with the recent eight-dimensional reformulation of ${\cal N}=4, 4D$ SYM theory in \cite{Sok}.

\noindent{\bf (f).} Applications in supergravity? Lacking the double-trace divergent structures in the
3-loop amplitude in ${\cal N}=(1,1)$ SYM theory is similar to the absence of 3-loop and 4-loop divergences for the
four-graviton amplitudes in ${\cal N}=4, 4D$ and ${\cal N}=5, 4D$ supergravities \cite{Bern:2012cd}, \cite{Tourkine:2012ip},
\cite{Bern:2012gh}, \cite{Bern:2014sna}. All these UV divergence cancelations could find a common explanation within
the harmonic superspace approach \footnote{For a recent relevant discussion see \cite{ASM}.}.

\ack
I thank the organizers of the conference ISQS'2016 for the kind hospitality in Prague.
 I am grateful to my  co-authors Guillaume Bossard, Andrei Smilga and  \fbox{Boris Zupnik}. A partial support from the RFBR grant no. 15-02-06670,
 grant of Russian Science Foundation no. 16-12-10306 and a grant of Heisenberg-Landau program is acknowledged.

\end{document}